\documentclass{PoS}
\usepackage{colordvi}
\usepackage{graphicx}
\usepackage{epsfig}
\usepackage{rotating}
\usepackage{bbm}

\def\bra#1#2{\ifx#2\ket\langle#1\else\langle#1\vert\fi#2}
\def\ket#1{\vert#1\rangle}

\newcommand{\Fig}{Fig.\ }
\newcommand{\Tab}{Tab.\ }
\newcommand{\Eq}{Eq.\ }

\leftline{\parbox{3cm}{\large\rm HU-EP-07/38 \\ ADP-07-13/T653}
\vspace{1cm}}

\title{Coulomb gauge studies of $SU(3)$ Yang-Mills theory on the lattice}
\ShortTitle{Coulomb gauge studies for $SU(3)$}

\author{\speaker{Aiko~Voigt}$^{\hspace{1mm}a,b}$, Ernst-Michael~Ilgenfritz$^{\,a}$, 
Michael~M\"uller-Preussker$^{\,a}$ and Andre~Sternbeck$^{\,c}$\\
~\\
$^{a}$ Humboldt Universit\"at zu Berlin, Institut f\"ur Physik, 12489 Berlin, Germany\\
$^{b}$ Max-Planck-Institut f\"ur Meteorologie, 20146 Hamburg, Germany\\
$^{c}$ CSSM, School of Chemistry \& Physics, University of Adelaide, SA 5005, Australia\\
E-mail: \email{aivoigt@physik.hu-berlin.de}, \email{aiko.voigt@zmaw.de},\\
\hspace*{1.15cm}\email{ilgenfri@physik.hu-berlin.de}, \email{mmp@physik.hu-berlin.de},\\
\hspace*{1.18cm}\email{andre.sternbeck@adelaide.edu.au}
}

\abstract{We study the infrared behaviour of lattice $SU(3)$ Yang-Mills theory 
in Coulomb gauge in terms of the ghost propagator, the Coulomb potential and the 
transversal and the time-time component of the equal-time gluon propagator. 
In particular, we focus on the Gribov problem and its impact on the observables. 
We observe that the simulated annealing method is advantageous for fixing the 
Coulomb gauge in large volumes. We study finite-size and discretization effects. 
While finite-size effects can be controlled by the cone cut, and the ghost 
propagator and the Coulomb potential become scaling functions with the cylinder cut,
the equal-time gluon propagator does not show scaling in the considered range 
of the inverse coupling constant. The ghost propagator is infrared enhanced. 
The Coulomb potential is now extended to considerably lower momenta and shows 
a more complicated infrared regime. The Coulomb string tension satisfies 
Zwanziger's inequality, but its estimate can be considered only preliminary 
because of the systematic Gribov effect that is particularly strong for the 
Coulomb potential.}

\FullConference{The XXV International Symposium on Lattice Field Theory\\
                 July 30-4 August 2007\\
                 Regensburg, Germany}

\begin{document}

\section{Motivation and Introduction}

According to the Gribov-Zwanziger scenario, confinement is related to the 
behaviour of certain gauge-variant two-point functions at small momenta. 
This is not only true for the Landau gauge. For the Coulomb gauge, similar 
predictions~\cite{Gribov:1977wm} exist:  
the transversal component $D_{tr}$ of the equal-time gluon 
propagator\footnote{Here and in the following, the norm of the spatial vector 
$\vec{q}=(q_1,q_2,q_3)$ is abbreviated by $q = |{\vec q}|$.}
\begin{equation}
 D^{\,ab}_{ij}({\vec q}) = \left< \tilde{A}^a_i(\vec{k}) \; \tilde{A}^b_j(-\vec{k}) \right> = \delta^{ab} \left(\delta_{\,ij}-\frac{q_{i}q_{j}}{\vec{q}^{2}} \right) D_{tr}(q^2)  
\end{equation}
should vanish in the limit $q \to 0$, whereas the ghost propagator $G$, defined by
\begin{equation}
G^{\,ab}({\vec q}) = \frac{1}{L^3}~ \sum_{{\vec x},\,{\vec y}}~\left< e^{\,i\,\vec{k}\cdot({\vec x} - {\vec y})} \left[M^{-1}\right]^{\,ab}_{{\vec x}\,{\vec y}} \right> = \delta^{ab}~G(q^2) \; , 
\end{equation}
is infrared enhanced\footnote{We use the shorthand notation $\vec{k}\cdot\vec{x} = \sum_{i=1}^{3} 2\pi k_{i}x_{i} / L_{i}$.}. The physical momentum is given by $q_i = (2/a) \cdot \sin(\pi k_i/L)$ with Fourier momenta $k_i=-L+1,...,L$.
$M$ denotes the Faddeev-Popov operator whose lattice version is given by
\begin{equation}
M^{\,ab}_{{\vec x}\,{\vec y}} = \sum_{i=1}^3 \mathfrak{Re}~\mathrm{Tr}~\left[\left\{T^a,T^b\right\}\left(U_{{\vec x},i} + U_{{\vec x}-\vec{i},i}\right)\delta_{\,{\vec x}\,{\vec y}} -2 T^b T^a \, U_{{\vec x},i} \, \delta_{\,{\vec x}+\vec{i},{\vec y}} -2 T^a T^b \, U_{{\vec x}-\vec{i},i} \, \delta_{\,{\vec x}-\vec{i},{\vec y}} \right] \; .
\end{equation}
Moreover, Zwanziger showed that the instantaneous Coulomb potential 
$V_{\rm coul}\left(r = \left| {\vec x}-{\vec y} \right|\right)$, 
which appears in the Coulomb gauge Hamiltonian through the elimination of 
longitudinal gluons and which is related to the Faddeev-Popov operator by
\begin{equation}
\delta^{\,ab}~V_{\rm coul}\left(q^2\right) = \frac{1}{L^3}~\sum_{{\vec x},\,{\vec y}}~\left< e^{\,i\,\vec{k}\cdot({\vec x}-{\vec y})} \left[M^{-1} \left(-\bigtriangleup\right) M^{-1}\right]^{ab}_{{\vec x}\,{\vec y}} \right> \; , 
\end{equation}
has to linearly rise with distance in real space.
This is a necessary condition to satisfy confinement usually defined through 
an area law for Euclidean Wilson loops. The latter translates to a linearly 
rising bound state potential $V_{\rm wilson}(r)$. More precisely, Zwanziger's 
inequality \cite{Zwanziger:2002sh} says that
$V_{\rm wilson}(r) \leq - \frac{4}{3}V_{\rm coul}(r)$
for large distances $r$. 
Zwanziger and Cucchieri pointed out that the time-time component of the 
equal-time gluon propagator $D^{\,ab}_{44}({\vec q}) =  \delta^{\,ab} D_{44}(q^2)$ 
is also related to the instantaneous Coulomb potential, i.e. 
$V_{\rm coul}(r) \simeq \left< A^a_4({\vec x}) \, A^a_4({\vec y}) \right>$ 
might serve as an useful estimate of the Coulomb potential \cite{Cucchieri:2000hv}. 
We have tested these statements by means of lattice calculations in 
$SU(3)$ Yang-Mills theory simulated with Wilson's plaquette action.
During this conference we learned about a similar investigation carried out by
Nakagawa et al.~\cite{NakagawaLat07}.

\section{Details of the simulation}

A summary of the lattices and parameters under investigation is given in 
\Tab \ref{tab:list_configs}. We used two different methods to fix the gauge. 
The overrelaxation algorithm with $\omega=1.70$ on small and $\omega=1.60$ 
on large lattices is compared with the simulated annealing method 
(with $T_{init}=0.45$ and $T_{final}=0.01$ as initial and final temperatures, 
1500 compound sweeps at linearly decreasing temperature, each consisting of one 
heatbath and three microcanonical sweeps, followed by obligatory finalizing 
overrelaxation). We investigated the Gribov effect by means of the 
first copy (fc) -- best copy (bc) strategy that was used in Landau gauge 
studies~\cite{Sternbeck:2005tk}. Since this approach is very demanding in terms 
of computing time, we have restricted the study of the Gribov problem to smaller 
lattices with $L^4 \leq 24^4$. On larger lattices only one gauge copy per 
configuration was generated by means of the simulated annealing algorithm. 
The bare ghost and equal-time gluon propagator are multiplicatively renormalized at a scale $\mu^2=9 {\rm~GeV}^2$. In the case of the Coulomb potential, we use an ultraviolet fit to \Eq \ref{eq:cp_ultraviolet} to fix the physical scale in y-direction.
 
\begin{table}[h]
\begin{center}
{\small
\begin{tabular}{|l c c c c c c r|}
\hline 
$L^4$   &   $\beta$   & $a^{-1}$ [GeV] & $a$ [fm] &$V$[fm$^4$]& \#conf &  $N_{cp}^{SA}$ & $N_{cp}^{OR}$ \\ \hline \hline
$12^4$ 	&   $5.8$     &   1.446        & 0.1364   &1.64$^4$  &   100  &  10              & 1	      \\
$16^4$ 	&   $5.8$     &   1.446        & 0.1364   &2.18$^4$  &   40   &  15              & 1	      \\
$24^4$ 	&   $5.8$     &   1.446        & 0.1364   &3.27$^4$  &   30   &  20              & 1	      \\
$32^4$ 	&   $5.8$     &   1.446        & 0.1364   &4.36$^4$  &   30   &  1               & -	      \\
$48^4$ 	&   $5.8$     &   1.446        & 0.1364   &6.55$^4$  &   20   &  1               & -	        \\
\hline
\hline
$12^4$ 	&   $6.0$     &   2.118        & 0.0932   &1.12$^4$  &   100  &  20              & 20	      \\     
$16^4$ 	&   $6.0$     &   2.118        & 0.0932   &1.49$^4$  &   60   &  30              & 30	      \\
$24^4$ 	&   $6.0$     &   2.118        & 0.0932   &2.24$^4$  &   40   &  40              & 40	      \\
$32^4$ 	&   $6.0$     &   2.118        & 0.0932   &2.98$^4$  &   30   &  1               & -	      \\
$48^4$ 	&   $6.0$     &   2.118        & 0.0932   &4.48$^4$  &   20   &  1               & -	        \\
\hline
\hline
$12^4$ 	&   $6.2$     &   2.914        & 0.0677   &0.81$^4$  &   100  &  10              & 1 	      \\ 
$16^4$ 	&   $6.2$     &   2.914        & 0.0677   &1.08$^4$  &   40   &  15              & 1	      \\
$24^4$ 	&   $6.2$     &   2.914        & 0.0677   &1.62$^4$  &   30   &  20              & 1	      \\
$32^4$ 	&   $6.2$     &   2.914        & 0.0677   &2.17$^4$  &   20   &  1               & -	        \\
\hline
\end{tabular} 
}
\caption{Ensembles and corresponding parameters as used in our investigation. 
$N_{cp}^{SA}$ and $N_{cp}^{OR}$ indicate the number of gauge copies 
per configuration generated with simulated annealing and overrelaxation, 
respectively.}
\label{tab:list_configs}
\end{center}
\end{table}
\vspace{-1.1cm}

\section{Gauge fixing}

Gauge fixing is performed by adjusting the gauge transformations $g({\vec x},t) \in SU(3)$ separately on each time-slice $t$ such that
\begin{equation}
F_U[g] = \frac{1}{3L^3} \; \sum_{{\vec x}} \; \sum_{i=1}^3 \mathfrak{Re}~\mathrm{Tr}~\left(\mathbbm{1} - \,^{g}U_{({\vec x},t),i} \right) \rightarrow\mbox{ minimum} \; .
\label{eq:Coulomb}
\end{equation}
The measurement of the $D_{44}$ propagator requires also to fix the remnant 
global gauge freedom with the temporal links involved, i.\ e.\ $U_{({\vec x},t),4} \to \tilde{g}_t~U_{({\vec x},t),4}~\tilde{g}_{t+1}$.  
\begin{figure}[ht]
\begin{center}
\hspace*{-0.6cm}\epsfig{figure=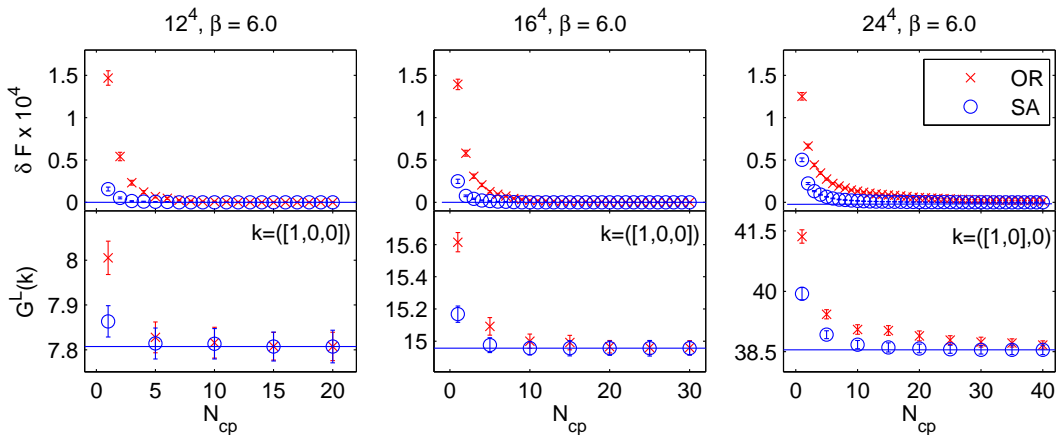,height=6cm}
\end{center}
\vspace{-0.7cm}
\caption{The Gribov ambiguity for different lattice sizes with OR and SA gauge 
fixing. Upper panels: The convergence of the relative difference 
$\delta F =1 - F^{cbc}/F^{bc}$ between the currently best functional value 
$F^{cbc}$ after $N_{cp}$ copies and the overall best value $F^{bc}$. Lower 
panels: the bare ghost propagator at the smallest lattice momentum 
measured for the currently best gauge copy after $N_{cp}$ copies.}
\label{fig:convergence_func_and_ghost}
\end{figure}
A possible Gribov effect on $D_{44}$ related to an ambiguity in fixing the 
remnant gauge freedom was not studied. Note also that minimizing $F_U[g]$ on 
the largest lattices was faced with severe critical slowing-down. About $5\%$ 
of the time-slices did not converge within 40000 (finalizing) overrelaxation 
steps. For these time-slices, the entire gauge fixing procedure was repeated 
until gauge fixing could be successfully accomplished.    

\section{Results}

\subsection{Gribov effect and its impact on the observables}

\begin{figure}[t]
\begin{center}
\hspace*{0.88cm}\epsfig{figure=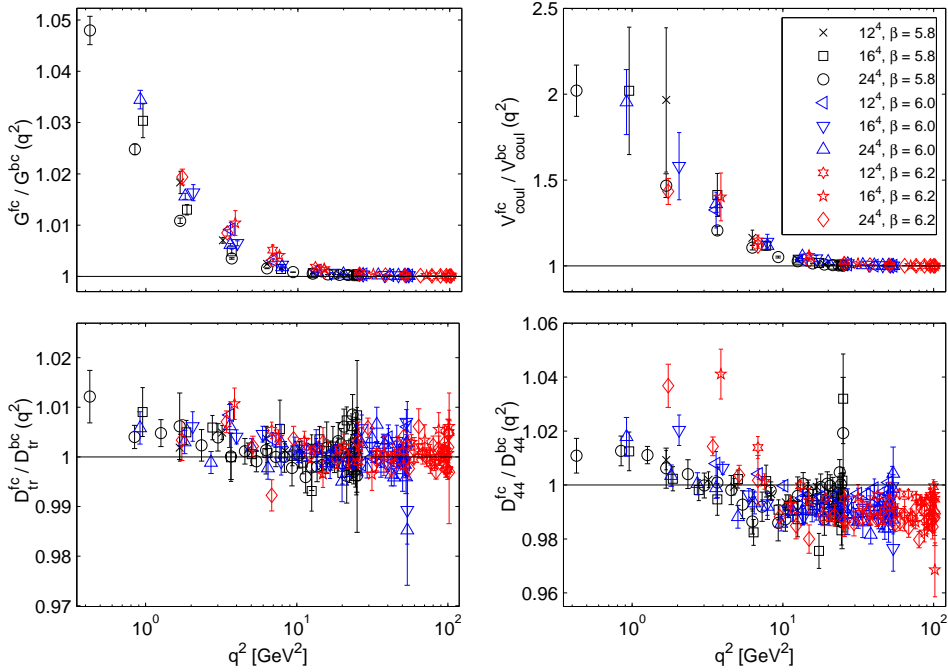,width=0.95\textwidth}
\end{center}
\vspace{-1cm}
\caption{Gribov effect on the propagators and the Coulomb potential: 
ratios of measured quantites as measured on the first OR copy and on the 
best of $N^{SA}_{cp}$ SA copies.}
\label{fig:obs_bcvsfc}
\end{figure}

Following the fc-bc strategy, we have first checked the number of gauge copies 
necessary to find a functional value $F$ close to the global maximum. This was 
done separately for both gauge fixing methods. For details of this strategy, 
we refer to analogous investigations in the Landau gauge~\cite{Sternbeck:2005tk}. 
The result is shown in \Fig \ref{fig:convergence_func_and_ghost} and points out 
that simulated annealing clearly needs less gauge copies to assure the convergence 
of the functional and hence of the observables. Therefore, simulated annealing 
should be the method of choice, in particular when going to larger lattices. 
\Fig \ref{fig:obs_bcvsfc} illustrates the impact of the Gribov effect by comparing 
the observables measured on the first overrelaxation gauge copy and on the best 
out of $N_{cp}^{SA}$ simulated annealing gauge copies. As already seen in 
Landau gauge, the infrared ghost propagator is systematically overestimated by 
about $5\%$. The Coulomb potential shows a similar behaviour, but now the first 
overrelaxation copy overestimates the Coulomb potential by up to $100\%$. 
The equal-time gluon propagator is less affected by the Gribov effect. 
Its transversal component shows a very weak ($1\%$) dependence on the Gribov 
copy in the infrared region. In contrast, the time-time component seems to be 
under-estimated systematically for large momenta. 

\subsection{Finite size and discretization effects}

\begin{figure}[ht]
\begin{center}
\epsfig{figure=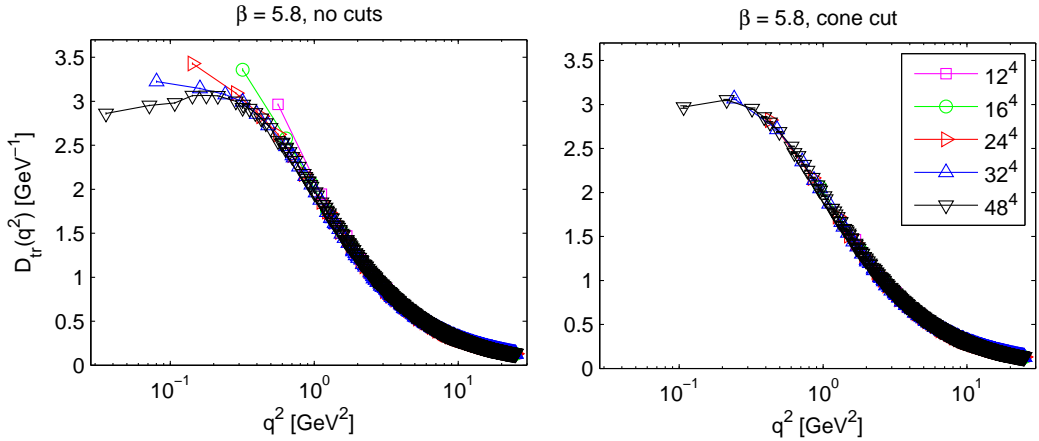,width=0.95\textwidth} 
\end{center}
\vspace{-0.5cm}
\caption{The finite-size effects (left) for the transversal gluon propagator, 
cured (right) by the cone cut.} 
\label{fig:gluondtrans_d_finitevolume_v3}
\end{figure}
\begin{figure}[hbt]
\begin{center}
\hspace*{0.1cm}
\epsfig{figure=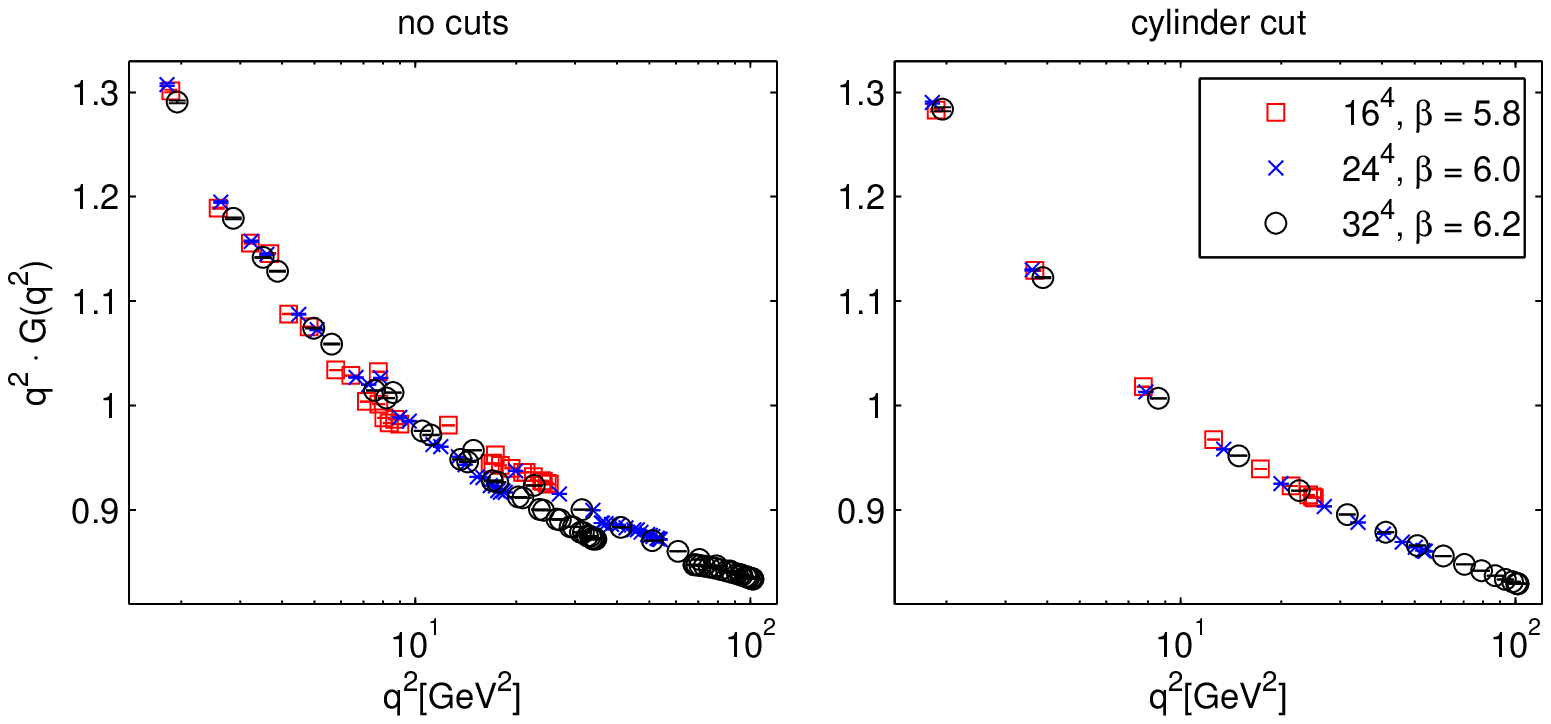,width=0.9\textwidth}
\end{center}
\vspace{-1cm}
\caption{Discretization effects (left) of the ghost dressing function, cured (right) 
by the cylinder cut.}
\label{fig:ghost_zg_errorinbeta_v3}
\end{figure}

We worked with a large number of lattice sizes and values of the inverse coupling 
$\beta$ in order to explore finite-size and discretization effects. We found that 
finite-size effects are visible but can be efficiently removed by the cone 
cut (see \cite{Sternbeck:2005tk} and references therein). 
\Fig \ref{fig:gluondtrans_d_finitevolume_v3} illustrates this for the transversal 
equal-time gluon propagator. Similarly, the cylinder cut minimizes discretization 
effects in the cases of the ghost propagator 
(see \Fig \ref{fig:ghost_zg_errorinbeta_v3}) and the Coulomb potential. 
More problematic is that we find the transversal and the time-time gluon 
propagator showing strong, systematic discretization effects. This has also been 
found by Nakagawa et al.~\cite{NakagawaLat07}.

\subsection{Infrared behaviour of the lattice observables}

\begin{figure}[htb]
\begin{center}
\epsfig{figure=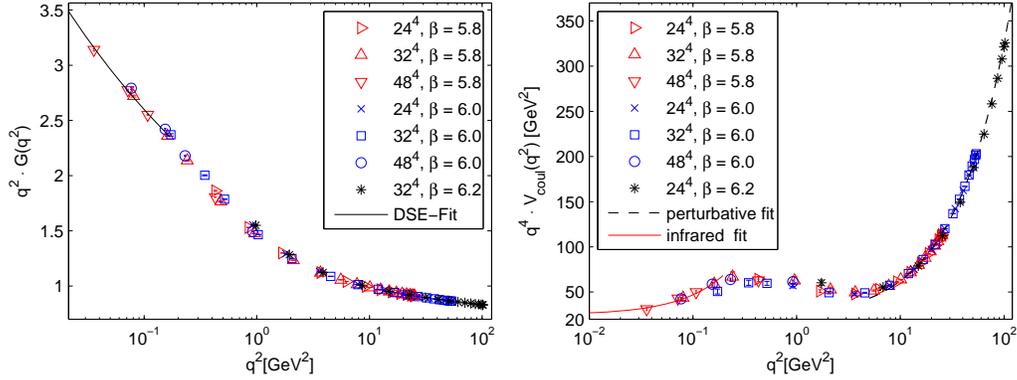,width=0.95\textwidth}
\end{center}
\vspace{-0.5cm}
\caption{Left: the ghost propagator diverging stronger than $1/q^2$ in the 
infrared, with a fit giving a too small infrared exponent. 
Right: the Fourier-transformed Coulomb potential times $q^4$, revealing the 
onset of a new infrared regime for $q^2 < 0.2 {\rm~GeV}^2$, with a fit 
giving $\sigma_{\rm coul}$.}
\label{fig:ghost_cp_final}
\end{figure}

All statements in this section rely on measurements on large lattices using a 
single (``first'') simulated annealing gauge copy and an appropriate combination 
of the cylinder and the cone cut. Since the equal-time gluon propagator does not 
show scaling for $\beta = 5.8$ ,..., $6.2$ we cannot give reliable statements 
concerning its behaviour in the infrared regime. However, the turnover of the 
transversal component for the smallest $\beta=5.8$ and $q^2 \leq 0.2 {\rm~GeV}^2$ 
is in agreement with a new infrared regime of the Coulomb potential observed for 
momenta below this value. The ghost propagator is infrared enhanced. An infrared 
fit of the ghost propagator to $q^{2}\cdot G(q^2) \simeq q^{-2\alpha}$ for momenta 
$q^2 \leq 0.17 {\rm~GeV}^2$ gives $\alpha = 0.19(2)$. This is by far too 
small compared to the predictions of DSE studies~\cite{Epple:2006hv} and, together 
with the high value of $\chi^{2}/\mbox{d.o.f.} \simeq 12$, indicates that the 
momentum region accessible in our study is not yet in the scope of the envisaged 
infrared power laws. We should also admit that our fit is clearly limited by the 
small number of points available in this momentum region. 

The Coulomb potential is infrared divergent. 
Plotting $q^4 \cdot V_{\rm coul}(q^2)$ reveals that a new infrared regime 
is opened for momenta smaller than $q^2 \sim 0.2 {\rm~GeV}^{2}$. 
After an intermediate plateau is reached, $q^4 \cdot V_{\rm coul}(q^2)$ 
decreases further with decreasing momenta. We used this momentum region to estimate 
the Coulomb string tension to $\sigma_{\rm coul}=(5.0 \pm 1.3) \sigma_{\rm wilson}$,
even higher than Zwanziger's inequality requires. To do so, first an ultraviolet fit 
\begin{equation}
\label{eq:cp_ultraviolet}
q^{4} \cdot V_{\rm coul}(q^2) = \frac{192 \pi^{2}}{121} \frac{q^{2}}{\ln(q^{2}/\Lambda^{2}_{\rm coul})} 
\hspace*{0.5cm} \mbox{~for~} q^2 \geq 20 {\rm~GeV}^2, \, \, \, \Lambda_{\rm coul} = 0.9(3) {\rm~GeV}
\end{equation}
and an infrared fit representing a sum of a constant force
and a $1/q^2$ Coulomb law, 
\begin{equation}
q^{4}\cdot V_{\rm coul}(q^2) = 8~\pi~\sigma_{\rm coul} + 4~\pi~C~q^2 
\hspace*{0.5cm} \mbox{~for~} q^2 \leq 0.16 {\rm~GeV}^2, \, \, \, C=17.8 \pm 5.9
\end{equation}
had to be performed. Note that our value of the Coulomb string tension does not take into 
account the strong Gribov effect on the Coulomb potential in the infrared.
It is strictly based on data for a single simulated annealing copy. Hence, the 
Coulomb string tension might be overestimated by a factor of two. 
This would be in better agreement with values obtained by measurements of partial 
Polyakov line correlators~\cite{Nakamura:2005ux}. We also stress that our result is in contrast to $SU(2)$ studies that had suggested
$\sigma_{\rm coul} \simeq \sigma_{\rm wilson}$ \cite{Cucchieri:2002su}. In view of the different results for the Coulomb potential and the time-time component of the gluon propagator, we conclude that the latter should not be used to estimate the Coulomb potential. The issue of non-scaling of the equal-time gluon propagator is currently investigated and  might be solved by a more advanced renormalization procedure.


\section*{Acknowledgements}

Part of this work is supported by DFG under contract FOR 465 
(Forschergruppe Gitter - Hadronen Ph\"anomenologie), and by the Australian Research Council. A major part of the simulations have been done on the IBM pSeries 690 at HLRN. We thank Hinnerk St\"uben for contributing parts of the code.

\end{document}